\documentclass[prd,twocolumn,aps,amsmath,amssymb,nofootinbib,preprintnumbers]
{revtex4}

\begin{document}
\newcommand{\drawsquare}[2]{\hbox{%
\rule{#2pt}{#1pt}\hskip-#2pt
\rule{#1pt}{#2pt}\hskip-#1pt
\rule[#1pt]{#1pt}{#2pt}}\rule[#1pt]{#2pt}{#2pt}\hskip-#2pt
\rule{#2pt}{#1pt}}

\newcommand{\Yfund}{\raisebox{-.5pt}{\drawsquare{6.5}{0.4}}}
\newcommand{\Yasymm}{\raisebox{-3.5pt}{\drawsquare{6.5}{0.4}}\hskip-6.9pt%
                     \raisebox{3pt}{\drawsquare{6.5}{0.4}}%
                    }
\newcommand{\Ysymm}{\Yfund\hskip-0.4pt%
                    \Yfund}
\def\symm{\Ysymm}
\def\bsymm{\overline{\Ysymm}}
\def\ls{\mathrel{\lower4pt\vbox{\lineskip=0pt\baselineskip=0pt
           \hbox{$<$}\hbox{$\sim$}}}}
\def\gs{\mathrel{\lower4pt\vbox{\lineskip=0pt\baselineskip=0pt
           \hbox{$>$}\hbox{$\sim$}}}}
\def\drawbox#1#2{\hrule height#2pt
        \hbox{\vrule width#2pt height#1pt \kern#1pt
              \vrule width#2pt}
              \hrule height#2pt}

\def\Fund#1#2{\vcenter{\vbox{\drawbox{#1}{#2}}}}
\def\Asym#1#2{\vcenter{\vbox{\drawbox{#1}{#2}
              \kern-#2pt       
              \drawbox{#1}{#2}}}}
\def\sym#1#2{\vcenter{\hbox{ \drawbox{#1}{#2} \drawbox{#1}{#2}    }}}
\def\fund{\Fund{6.4}{0.3}}
\def\asymm{\Asym{6.4}{0.3}}
\def\bfund{\overline{\fund}}
\def\basymm{\overline{\asymm}}


\newcommand{\beq}{\begin{equation}}
\newcommand{\eeq}{\end{equation}}
\def\ls{\mathrel{\lower4pt\vbox{\lineskip=0pt\baselineskip=0pt
           \hbox{$<$}\hbox{$\sim$}}}}
\def\gs{\mathrel{\lower4pt\vbox{\lineskip=0pt\baselineskip=0pt
\def\lsim{\mathrel{\lower4pt\vbox{\lineskip=0pt\baselineskip=0pt
           \hbox{$<$}\hbox{$\sim$}}}}
\def\gsim{\mathrel{\lower4pt\vbox{\lineskip=0pt\baselineskip=0pt
           \hbox{$>$}\hbox{$\sim$}}}}           \hbox{$>$}\hbox{$\sim$}}}}

\title{Reheating in supersymmetric high scale inflation}

\author{Rouzbeh Allahverdi$^{1,2}$}
\author{Anupam Mazumdar$^{3}$}
\affiliation{$^{1}$~Perimeter Institute for Theoretical Physics, Waterloo, ON,
N2L 2Y5, Canada. \\
$^{2}$~Department of Physics and Astronomy, McMaster University, Hamilton,
ON, L8S 4M1, Canada. \\
$^{3}$~NORDITA, Blegdamsvej-17, Copenhagen-2100, Denmark.}

\begin{abstract}
Motivated by Refs~\cite{am1,am2}, we analyze how the inflaton decay
reheats the Universe within supersymmetry. In a non-supersymmetric
case the inflaton usually decays via preheating unless its couplings to other
fields are very small.
Naively one would expect that supersymmetry enhances bosonic
preheating as it introduces new scalars such as squarks and sleptons.
On the contrary, we point out that preheating is unlikely within
supersymmetry. The reason is that flat
directions in the scalar potential, classified by gauge
invariant combinations of slepton and squark fields, are generically
displaced towards a large vacuum expectation value (VEV) in the early
Universe. They induce supersymmetry preserving masses to the
inflaton decay products through the Standard Model Yukawa couplings,
which kinematically blocks preheating for VEVs $> 10^{13}$ GeV.  The decay
will become allowed
only after the flat directions start oscillating, and once the flat
direction VEV is sufficiently redshifted. For models with weak scale
supersymmetry, this generically happens at a Hubble expansion rate: $H
\simeq \left(10^{-3}-10^{-1}\right)~{\rm TeV}$, at which time the
inflaton decays in the perturbative regime.  This is to our knowledge
first analysis where the inflaton decay to the Standard Model
particles is treated properly within supersymmetry.  There are number
of important consequences: no overproduction of dangerous
supersymmetric relics (particularly gravitinos), no resonant
excitation of superheavy dark matter, and no non-thermal leptogenesis
through non-perturbative creation of the right-handed
(s)neutrinos. Finally supersymmetric flat directions can even spoil
hybrid inflation all together by not allowing the auxiliary field
become tachyonic.
\end{abstract}
\preprint{NORDITA-2006-5}
\maketitle

\section{Introduction}

Reheating after inflation connects the observable sector to the
inflaton sector, which does not apriori carry Standard Model (SM)
charges.  Often inflationary paradigm is realized with a {\it SM gauge
singlet} inflaton whose origin and couplings to the matter cannot be
explained within SM or its minimal extensions~\cite{Lyth}.  Therefore
transferring inflaton energy density into the SM degrees of freedom is
the single most important phenomenon which would guarantee a
successful big bang nucleosynthesis (BBN)~\cite{BBN}.

The inflaton decay is the first and the most relevant part of
reheating.  Only one-particle decay of the non-relativistic inflaton
quanta were considered initially~\cite{reheat}.  The treatment is
valid if the energy transfer to the fields which are coupled to the
inflaton takes place over many inflaton oscillations.  This requires
that the inflaton couplings to the SM fields are sufficiently small,
which is justifiable if the inflaton (being a gauge singlet) couples
through non-renormalizable operators to the matter sector.  Usually it
is assumed that the plasma reaches complete kinetic and chemical
equilibrium immediately after all the inflaton quanta have decayed,
which is true in a non-supersymmetric
scenarios~\cite{Thermalization,Jaikumar,am2}.

It was also pointed out that the coherent oscillations of the inflaton
can create particles non-perturbatively~\cite{preheat1,preheat2}. This
mechanism is called preheating and it is particularly efficient when
the final products are bosonic degrees of freedom. It only takes about
two dozens of oscillations to transfer the energy from the homogeneous
condensate to non-zero modes of the final state(s)~\cite{preheat2}.
However, despite efficiently transferring the energy, preheating does
not result in a {\it complete} decay of the inflaton. In some cases
the inflaton condensate fragments to form non-topological solitons
which decays through surface
evaporation~\cite{Enqvist-Soliton}. Irrespective of theses situations
it is very challenging to understand thermalization of the preheated
plasma.

Although an epoch of perturbative reheating is an essential ingredient
of any potentially realistic cosmological model~\cite{Jed}, preheating
remains a possibility which can give rise to rich physical phenomena
ranging from non-thermal production of
particles~\cite{preheat1,preheat2}, production of
gravitinos~\cite{Maroto,non-pert2,non-pert3} and moduli~\cite{Moduli},
large isocurvature perturbations~\cite{Robert-Iso}, amplifying gravity
waves~\cite{Igor}, generation of large
non-Gaussianity~\cite{Enqvist-NG,Asko-ng}, non-thermal source for
leptogenesis and baryogenesis~\cite{Peloso}, and the formation of
topological and non-topological
solitons~\cite{Tkachev,Enqvist-Soliton}.

Supersymmetry (SUSY) provides the framework for most widely studied
extensions of the SM physics beyond the electroweak scale.  Therefore,
given the large energy density of the inflaton, it is then pertinent
to ask how the inflaton decays within SUSY. Particularly, into the
relativistic degrees of freedom of the minimal supersymmetric SM
(MSSM). Then the next question is how quickly do they thermalize.
This very last question was recently addressed in Ref.~\cite{am2}.

In a Non-SUSY case it is known that the inflaton decay products
thermalize very quickly because of the efficiency of interactions
mediated by the {\it massless gauge bosons} of the
SM~\cite{Thermalization}.  Therefore the reheat temperature is mainly
governed by the inflaton decay width.  In Refs.~\cite{am1,am2}, we
pointed out that the process of thermalization is in general
painstakingly slow within SUSY. The Universe loiters in a phase of a
quasi-thermal equilibrium after the decay of the inflaton, which
results in a very low reheat temperature, i.e. $T_{R} \sim {\cal
O}(10^{3}-10^{7})$~GeV. The final reheat temperature does not depend
on the decay width of the inflaton rather on a thermalization time
scale which depends on the vacuum expectation value (VEV) of the
squarks and sleptons.

What was not addressed in Ref.~\cite{am2} is the initial phase of the
inflaton decay. In this paper we will fill up that gap. We would like
to know how the inflaton decays: whether {\it perturbatively} or {\it
non perturbatively}.  This simple question is so relevant that
depending on the nature of the inflaton decay there would be different
consequences all together. For instance production of baryons, cold
dark matter, magneto-genesis, electroweak baryogenesis, production of
dangerous relics and their abundances depend on the nature of the
primordial plasma. Therefore this is an important topic which relates
the early Universe physics to phenomenology.

An important fact is the presence of flat directions along which the
scalar potential identically vanishes in the limit of exact SUSY. In
the MSSM alone there are nearly $300$ flat directions~\cite{gkm},
which are made up of {\it gauge invariant} combinations of squarks,
sleptons and Higgses. These are none but the simplest examples of
moduli near points of enhanced symmetry. During inflation more than
one MSSM flat directions (orthogonal in flavor basis)~\cite{ASKO} are
expected to develop large VEVS, for a review, see~\cite{Enqvist-REV}.

A large VEV of the MSSM flat directions, during and after inflation,
spontaneously breaks the SM gauge group, and gives masses to the gauge
bosons and gauginos similar to the Higgs mechanism. Many of the flat
directions break the entire SM gauge group~\cite{ejm}. The flat
direction VEV also induces large SUSY preserving masses to (s)quarks,
(s)leptons and Higgs/Higgsino fields during and after inflation. As we will
see, such
large masses kinematically prohibit non-perturbative inflaton decay
into MSSM fields. The initial stage of inflaton oscillations thus
produces no significant fraction of MSSM particles~\cite{am2}. The
decay will occur much later after the flat direction oscillations have
started and their VEV has been sufficiently redshifted. Such a delayed
decay of the inflaton is typically in the perturbative regime.

Rest of the paper is organized as follows. In Section II we briefly
discuss inflaton decay in a non-SUSY case. We highlight inflaton
couplings to the MSSM fields in Section III, and MSSM flat direction
couplings to the inflaton decay products in Section IV. We then
discuss the flat direction dynamics during and after inflation in
Section V. In Section VI we explain how the VEV of flat directions
prevent non-perturbative decay of the inflaton.  We illustrate in
Section VII how the inflaton eventually decays perturbatively, and
discuss various cosmological consequences in Section VIII.  Finally we
briefly mention our conclusion.  We have added appendices discussing
some minute details for the paper to be self-contained.

\section{Inflaton decay in a non-SUSY case}
\label{BRP}

First let us briefly review the initial stage of the inflaton decay
which is typically non-perturbative, i.e. preheating, in a non-SUSY
case.  Our focus is on bosonic preheating which acts most
efficiently in transferring the energy density from the inflaton
oscillations. We consider models of large field inflation, such as
chaotic inflation, for which bosonic preheating is most pronounced.
The relevant renormalizable couplings between the inflaton $\phi$
and a scalar field $\chi$ will read from the following potential:
\beq \label{nonpot}
V = {1 \over 2} m^2_{\phi} \phi^2 + \sigma \phi \chi^2 + h^2 \phi^2 \chi^2 +
\lambda \chi^4\,, \eeq
where we have considered $\phi$ and $\chi$ to be real. Here $\sigma$
is a coupling which has a [mass] dimension. The only scalar field in
the SM is the Higgs doublet. Therefore in a realistic case $\chi$
denotes the real and imaginary parts of the Higgs
components~\footnote{Since the SM fermions are chiral, the inflaton
can only couple to them through dimension-$5$ operators. The same
holds for coupling to gauge bosons where the inflaton is coupled to
gauge field strengths. Such couplings are non-renormalizable and
suppressed compared to those in Eq.~(\ref{nonpot}), and hence
negligible (we consider the scale of non-renormalizable operator is
governed by $M_{\rm P}$).}.  Note that the cubic interaction term is
required for a complete inflaton decay.  The quartic self-coupling of
$\chi$ is required to bound the potential from below along the $\chi$
direction. The dimensionless couplings $\sigma/m_{\phi}$ and $h$ (as
well as $\lambda$) are not related to each other, hence either of the
cubic or the quartic terms can dominate at the beginning of inflaton
oscillations (i.e. when the Hubble expansion rate is $H(t) \simeq
m_{\phi}$ and the amplitude of oscillations is ${\hat \phi} \sim {\cal
O}(M_{\rm P})$).

\begin{itemize}

\item{$\sigma \ll h^2 M_{\rm P}$: In this regime the $h^2 \phi^2
\chi^2$ term is dominant at the beginning of the inflaton
oscillations.  This case has been studied in detail in first two
references of~\cite{preheat2}.  For a nominal value of the inflaton
mass, $m_{\phi} = 10^{13}$ GeV, non-perturbative $\chi$ production
with a physical momentum, $k \ls \left( h m_{\phi} {\hat
\phi}\right)^{1/2}$, takes place if $h > 10^{-6}$.  Particle
production is particularly efficient if $h > 3 \times 10^{-4}$, and
results in an explosive transfer of energy to $\chi$ quanta which ends
when re-scatterings destroy the inflaton condensate~\footnote{Further
note that the $h^2 \phi^2 \chi^2$ term does not produce any significant
non-Gaussianity if $h > 10^{-5}$, see the first reference
of~\cite{Enqvist-NG}.}.  The whole process happens over a time scale
$\sim 150 m^{-1}_{\phi}$, which depends logarithmically on $h$
~\footnote{In a non-SUSY case efficient
preheating happens over a narrow window $3 \times 10^{-4} \leq h \leq
10^{-3}$. The reason is that the $h^2 \phi^2 \chi^2$ term yields a
quartic self-coupling for the inflaton at a one-loop level which is
constrained by the COBE normalization of the density
perturbations~\cite{Robert-REV,Enqvist-REV}. However in SUSY this
correction is canceled out by that from fermionic partner of $\chi$,
so in principle one could expect a rather broader range of parameter
space within SUSY.}.}

\item{$\sigma \gg h^2 M_{\rm P}$: In this regime the cubic term
$\sigma \phi \chi^2$ dominates.  This case was recently considered in
Refs.~\cite{Natalia,dfkpp}, where the the $\chi$ field becomes
tachyonic during half of each oscillation.  For $\sigma >
m^2_{\phi}/M_{\rm P}$ (which amounts to $\sigma > 10^7$ GeV for
$m_{\phi} = 10^{13}$ GeV) this tachyonic instability transfers energy
from the oscillating condensate very efficiently to the $\chi$ quanta
with a physical momentum $k \ls \left(\sigma {\hat \phi}
\right)^{1/2}$. Particle production ceases when the back-reaction from
$\chi$ self-coupling induces a mass-squared $\gs \sigma {\hat
\phi}$. Depending on the size of $\lambda$, most of the energy density
may or may not be in $\chi$ quanta by the time backreaction becomes
important~\cite{dfkpp}~\footnote{In Ref.~\cite{dfkpp}, the authors
attempted to motivate the cubic coupling from SUSY. However they
missed vital ingredients which exist in a realistic case, such as the strength
of the SM couplings,
contributions from SUSY flat directions, etc. In this paper we wish to
note that neither the couplings nor the interactions are taken
arbitrarily.}.}

\end{itemize}

Couple of points to note here. In the borderline regime $ \sigma
\sim h^2 M_{\rm P}$, the cubic and quartic interaction terms are
comparable.  The inflaton decay happens due to a combination of
resonant and tachyonic instabilities. If $h \ll m_{\phi}/M_{\rm P}$
and $ \sigma \ll m^2_{\phi}/M_{\rm P}$, the inflaton decays
perturbatively via the cubic interaction term. However this requires
very small couplings: $h,\left(\sigma/m_{\phi}\right) < 10^{-6}$.
Therefore, unless the inflaton is only gravitationally coupled to
other fields, the initial stage of its decay will be generically
non-perturbative.

The plasma from the non-perturbative inflaton decay eventually reaches
full thermal equilibrium, though, at time scales much longer than that
of preheating itself~\cite{fk,mt,kp}. The occupation number of
particles is $f_{k}\gg 1$ in the meantime. This implies that dangerous
relics (such as gravitino and moduli) can be produced much more
copiously in the aftermath of preheating than in full thermal
equilibrium~\cite{Moduli,am1,kp}. This is a negative aspect of an
initial stage of preheating. One usually seeks a late stage of entropy
release, in order to dilute the excess of relics. As we shall show,
supersymmetry naturally provides us a tool to undo preheating completely.

\section{Inflaton couplings to matter in SUSY}
\label{ICM}

Inflaton couplings to (MS)SM fields is of utmost importance for
(p)reheating. In all the relevant papers, for instance see
Refs.~\cite{preheat1,preheat2}, inflaton couplings to matter has not
been dealt with carefully. Only toy models have been considered which
have no relevance to SM physics~\footnote{Fermionic preheating has
been discussed in Refs.~\cite{Kofman}, but the importance of SM gauge
invariance was grossly neglected. In fact the inflaton couples to SM
fermions through non-renormalizable dimension $5$ operators, and
therefore preheating into SM fermion is unlikely.}. The importance of
gauge invariance was first highlighted in
Refs.~\cite{Jaikumar,abm,am2,marieke}~\footnote{In Refs.~\cite{Asm}
the inflaton belonged to a gauge sector which can carry SM charges
based on the ideas of {\it assisted inflation}~\cite{Liddle}. In which
case the inflaton couplings to matter are governed by the usual
Yukawas. However it is hard to construct realistic models without
gauge singlets.}.

In almost all known $F$ and/or $D$-term models of inflation the
inflaton, $\phi$, is considered to be an absolute gauge singlet. Then
the main question arises how the inflaton couples to the matter. This
is one of the most pertinent issues which connects inflation to a hot
big bang cosmology.

First note the field content of MSSM which is governed by the
following superpotential:
\begin{equation}
\label{mssm}
W_{\rm MSSM}=\lambda_u {\bf Q} {\bf H}_u {\bf u} + \lambda_d {\bf Q} {\bf H}_d
{\bf d} + \lambda_e {\bf L} {\bf H}_d {\bf e}~ + \mu {\bf H}_u {\bf H}_d\,,
\end{equation}
where ${\bf H}_u, {\bf H}_d, {\bf Q}, {\bf L}, {\bf u}, {\bf d},
{\bf e}$ in Eq.~(\ref{mssm}) are chiral superfields representing the
two Higgs fields (and their Higgsino partners), left-handed (LH)
(s)quark doublets, right-handed (RH) up- and down-type (s)quarks, LH
(s)lepton doublets and RH (s)leptons respectively.  The
dimensionless Yukawa couplings $\lambda_{u}, \lambda_{d},
\lambda_{e}$ are $3\times 3$ matrices in the flavor space, and we
have omitted the gauge and flavor indices. The last term is the
$\mu$ term, which is a supersymmetric version of the SM Higgs boson
mass.

There exist two gauge-invariant combinations of only two superfields:
\beq \label{two}
{\bf H_u} {\bf H_d} ~ ~ , ~ ~ {\bf H_u} {\bf L}.
\eeq
The combinations which include three superfields are:
\beq \label{three}
{\bf H_u} {\bf Q} {\bf u} ~ , ~ {\bf H_d} {\bf Q} {\bf d} ~ , ~ {\bf H_d}
{\bf L} {\bf e} ~, ~ {\bf Q} {\bf L}
{\bf d} ~ , ~ {\bf u} {\bf d} {\bf d} ~ , ~ {\bf L} {\bf L} {\bf e}.
\eeq
SUSY together with gauge symmetry requires that the inflaton
superfield be coupled to these combinations~\footnote{It is possible
that the inflaton mainly decays to another singlet (for example, the
RH neutrinos) superfield, see the discussion in
section~\ref{ICSGS}}. The terms ${\bf \Phi} {\bf H_u} {H_d}$ and ${\bf
\Phi} {\bf H_u} {\bf L}$ have dimension four, and hence are
renormalizable. On the other hand, the interaction terms that couple
the inflaton to the combinations in Eq.~(\ref{three}) have dimension
five and are non-renormalizable. In following we focus on
renormalizable interactions of the inflaton with matter which play the
dominant role in its decay~\footnote{We note that terms representing
gauge-invariant coupling of the inflaton to the gauge fields and
gauginos are also of dimension five, and hence preheating into
them will be suppressed.}.

\subsection{Two choices of renormalizable couplings}

The simplest case is when the inflaton is coupled to matter via
superpotential terms of the form:
\begin{eqnarray} \label{rensup}
2 g {\bf \Phi} {\bf H_u} {\bf H_d} ~ ~ \,,~ ~
2 g {\bf \Phi} {\bf H_u} {\bf L} \, .
\end{eqnarray}
where $g$ can be as large as ${\cal O}(1)$. The factor of $2$ as we
shall see, leads to convenience in field redefinitions. Besides the SM
gauge group the MSSM Lagrangian is also invariant under a discrete
$Z_2$ symmetry namely ``$R$-parity''. This symmetry assigns the number
$R = {(-1)}^{3B + L + 2S}$ to the component fields where $B,~L,~S$ denote
the baryon number, lepton number and spin respectively. This amounts
to $+1$ for the SM fields and $-1$ for their supersymmetric
partners. As a result the lightest supersymmetric particle (LSP) will
be stable and can account for dark matter in the Universe. This is one
of the most remarkable cosmological features of MSSM.

\subsection{Preserving $R$-parity}

Preserving $R$-parity at the renormalizable level further constrains
inflaton couplings to matter. Note that ${\bf H}_u {\bf H}_d$ is
assigned $+1$ under $R$-parity, while ${\bf H}_u {\bf L}$ has the
opposite assignment $-1$. Therefore only one of the couplings in
Eq.~(\ref{rensup}) preserves $R$-parity: ${\bf \Phi} {\bf H}_u {\bf
H}_d$ if
$R_{\bf \Phi} = +1$,
and ${\bf \Phi} {\bf H}_u {\bf L}$
if
$R_{\bf \Phi} = -1$ (such as models where the RH sneutrino plays the
role of the inflaton~\cite{sninfl}). Therefore the renormalizable
inflaton coupling to matter can be represented as
\beq \label{infsup}
2 g {\bf \Phi} {\bf H}_u {\bf \Psi} \,
\eeq
where
\begin{eqnarray} \label{def1}
{\bf \Psi} = {\bf H}_u ~ ~ ~ ~ {\rm if} ~ R_{\bf \Phi} = +1 \, , \nonumber \\
{\bf \Psi} = {\bf L} ~ ~ ~ ~ {\rm if} ~ R_{\bf \Phi} = -1 \,.
\end{eqnarray}
Taking into account of the inflaton superpotential mass term:
$\left(m_{\phi}/2 \right){\bf \Phi} {\bf \Phi}$, and after defining
\begin{eqnarray} \label{def2}
{\bf X}_{1,2} = {\left({\bf H}_u \pm {\bf \Psi} \right) \over \sqrt{2}}
\, ,
\end{eqnarray}
and with the help of Eqs.~(\ref{fplusd},\ref{fddefs}), see
appendix~\ref{BRM}, we find the {\it renormalizable part of the
potential} which is relevant for the inflaton decay into MSSM scalars
is given by:
\begin{eqnarray}
\label{infpot} V \supset {1 \over 2} m^2_{\phi} {\phi}^2 + g^2
{\phi}^2 \chi^2 \pm
{1 \over \sqrt{2}} g m_{\phi} \phi \chi^2 \,,
\end{eqnarray}
where $\chi$ denotes the scalar component of ${\bf X}_{1,2}$
superfields, and we have only considered the real parts of the
inflaton, $\phi$, and $\chi$ field. Further note that the cubic
interaction term appears with different signs for $\chi_1$ and
$\chi_2$, but this is irrelevant during inflaton oscillations. We have
neglected the inflaton coupling to the fermionic partners of $\chi$ as
we focus on the bosonic preheating here. However our analysis will
follow similarly to the fermionic case and the same conclusions hold
for fermionic preheating too.

In addition to the terms in Eq.~(\ref{infpot}) there are also the
self- and-cross-couplings, $\left(g^2/4\right) {\left(\chi^2_1 -
\chi^2_2 \right)}^2 + \alpha \chi^2_1 \chi^2_2$, arising from the
superpotential and $D$-terms respectively ($\alpha$ is a gauge fine
structure constant). Therefore even in the simplest SUSY set up the
scalar potential is more involved than the non-SUSY case given in
Eq.~(\ref{nonpot}), which can alter the picture of preheating
presented in the literature~\cite{ac,pr}. Note however that these
terms become important {\it after} particle production has started. Here
we focus on the terms in Eq.~(\ref{infpot}) which are relevant for
particle creation {\it from the very beginning of the oscillations}.

A remarkable feature in Eq.~(\ref{infpot}) is that SUSY naturally
relates the strength of cubic $\phi \chi^2$ and quartic $\phi^2 \chi^2$
interactions. {\it We re-emphasize that the cubic term is required for
complete decay of the inflaton field}. This is a natural consequence
of SUSY which holds so long as the inflaton mass is larger than the
soft SUSY breaking masses.



\section{Flat direction couplings to inflaton decay products}
\label{FDCIDP}

Consider a MSSM flat direction, $\varphi$, with the corresponding
superfield denoted by $\varphi$ (only for flat directions we are
denoting the superfield and the field with the same notation). For a
brief discussion on MSSM flat directions, see
Appendices~\ref{BRM},~\ref{append}. Note that the ${\varphi}$ and ${\bf
X}$ superfields are linear combinations of the MSSM superfields, see
Eq.~(\ref{def2}), and hence are coupled through the MSSM
superpotential in Eq.~(\ref{mssm}). The couplings are nothing but the
(MS)SM Yukawas. Then the MSSM superpotential can be recast in the
following form:
\beq\label{rel}
W \supset \lambda_1 {\bf H}_u {\varphi}
{\Sigma}_1 + \lambda_2 {\bf \Psi} {\varphi} {\Sigma}_2 + ... ,
\eeq
where ${\Sigma}_{1,2}$ are some MSSM superfields~\footnote{Note that
${\Sigma}_1 \neq {\bf \Psi}$ and ${\bf \Sigma}_2 \neq {\bf H}_u$, since
${\varphi}$ is a non-gauge-singlet.}. For example consider the case
where $\varphi$ is a flat direction classified by the ${\bf u} {\bf d}
{\bf d}$ monomial, and ${\bf \Psi} = {\bf H}_d$. In this case ${\bf
\Sigma}_{1,2}$ are ${\bf Q}$ superfields and $\lambda_{1,2}$
correspond to ${\lambda}_u$ and $\lambda_d$ respectively, see
Eq.~(\ref{mssm}).  After using Eq.~(\ref{def2}) we find:
\beq \label{relsup} W \supset {\lambda_1 \over \sqrt{2}} {\bf X}
{\varphi} {\Sigma_1} + {\lambda_2 \over \sqrt{2}} {\bf X} {\varphi}
{\Sigma_2}\,. \eeq
This results in:
\beq \label{chimass}
V \supset \lambda^2 {\vert \varphi \vert}^2 {\chi^2} ~ ~ , ~ ~
\lambda \equiv \left({\lambda^2_1 + \lambda^2_2 \over 8}\right)^{1/2}\,,
\eeq
where we have again considered the real part of $\chi$.

Let us determine the strength of flat direction coupling to $\chi$,
which is denoted in Eq.~(\ref{chimass}) by $\lambda$. Note that the
first generation of (s)leptons and (s)quarks have a Yukawa coupling
$\sim {\cal O} (10^{-6}-10^{-5})$, while the rest of the SM Yukawa
couplings are $\geq 10^{-3}$.  The MSSM flat directions can be grouped
in $6$ categories mentioned in Appendix~\ref{append}, out of which
\begin{itemize}
\item{Only 11 directions: $3~{\bf udd}$s,~$6~{\bf QdL}$s, $1~{\bf
LLddd}$ and $1~{\bf LLe}$ have couplings to MSSM particles/sparticles
such that $\lambda < {\cal O}(10^{-5})$,}
\item{Rest of the flat directions have $\lambda \geq 3 \times
10^{-4}$.}
\end{itemize}
%

\section{Flat direction potential}
\label{FDDAI}

The flat directions are lifted by soft SUSY breaking mass term,
$m_{0}\sim {\cal O}({\rm TeV})$, Hubble induced corrections and
superpotential corrections of type: $W \sim \lambda_{n}{{\varphi}^n}/n
{M^{n-3}}$~\cite{drt}
\beq \label{nonren}
V \supset \left(m^2_0 + c_H H^2 \right) {\vert \varphi \vert}^2 +
\lambda^2_{n} {{\vert \varphi \vert}^{2(n-1)} \over M^{2(n-3)}}\,,
\eeq
with $n \geq 4$. Here $M$ is the scale of new physics which induces
the non-renormalizable terms, typically the Planck scale $M = M_{\rm
P}$ or the grand unification scale $M = M_{\rm GUT}$.

Note that $c_H$ can have either sign. If $c_H \gs 1$, the flat
direction mass is $> H$. It therefore settles at the origin during
inflation and remains there~\footnote{This has a similar origin as a
supergravity inflationary $\eta$-problem, see~\cite{Lyth}.} . Since
$\langle \varphi \rangle = 0$ at all times, the flat direction will
have no interesting consequences in this case. However there is a
large class of theories which predicts $c_H <0$ and also $c_H\ll
+1$. Negative $c_H$ may arise naturally if the inflaton and MSSM
flat directions have positive higher order couplings in the K\"ahler
potential, i.e. $\Phi^{\dagger}\Phi
\varphi^{\dagger}\varphi$~\cite{drt}, such that all the eigenvalues
of the K\"ahler matrix is positive definite. There is {\it no
symmetry} which prohibits such couplings.

Moreover string theory, which we believe will provide the true low
energy effective theory, also generically predicts no-scale type
K\"ahler potential based on Heisenberg symmetry~\cite{NO-Scale}, which
at tree level gives no correction to the flat direction mass,
i.e. $c_H=0$. However a Hubble induced mass term is generated at a
loop level, because MSSM superpotential (as well as $D$-terms) breaks
the Heisenberg symmetry, which induces a calculable but small
contribution, i.e. $c_H \leq 10^{-2}$~\cite{gmo}. Moreover, even
starting with $c_H > 0$ at a high scale, it is possible that $c_H$
quickly changes sign due to loop corrections from large Hubble-induced
SUSY breaking terms~\cite{adm}.

In the absence of $c_H \gs 1$, the flat direction remains flat during
inflation as the Hubble expansion rate is $H_{\rm I} \gg
m_0$. Therefore quantum fluctuations are free to accumulate (in a
coherent state) along $\varphi$ and form a condensate with a large
VEV, $\varphi_0$.  Because inflation smoothes out all gradients, only
the homogeneous condensate mode survives. However, the zero point
fluctuations of the condensate impart a small, and in inflationary
models a calculable, spectrum of perturbations on the
condensate~\cite{Enqvist-REV}.

If the higher-order superpotential term is forbidden, due to an
$R-$symmetry (or a set of $R-$symmetries)~\cite{dis}, then we
naturally have, $\varphi_0 \sim M_{\rm P}$~\cite{drt}.  On the other
hand, $\varphi_0 \ll M_{\rm P}$ will be possible if non-renormalizable
superpotential terms are allowed.  As shown in~\cite{gkm}, in the
absence of any $R$-symmetry, all the MSSM flat directions are lifted
by higher-order terms with $n \leq 9$. If a flat direction is lifted
at the superpotential level $n$, the VEV that it acquires during
inflation cannot exceed:
\begin{equation}
\varphi_{0} \sim \left(H_{\rm I}M^{n-3}\right)^{1/(n-2)}\,,
\label{ultivev}
\end{equation}
where $H_{\rm I}$ is the expansion rate of the Universe in the
inflationary epoch.

After inflation, $H(t) \propto t^{-1}$, the flat direction stays at
a relatively larger VEV due to large Hubble friction term, note that
the Hubble expansion rate gradually decreases but it is still large
compared to $m_{0}$. When $H(t) \simeq m_0$, the condensate along
the flat direction starts oscillating around the origin with an
initial amplitude $\varphi_{\rm in} \sim \left(m_0 M^{n-3}_{\rm
P}\right)^{1/(n-2)}$. From then on $\langle \varphi \rangle$ is
redshifted by the Hubble expansion $\propto H$ for matter dominated
and $\propto H^{3/4}$ for radiation dominated Universe.

\section{No preheating in SUSY}
\label{KBP}

In order to understand the preheating dynamics it is important to take
into account of $\chi$ coupling to the inflaton $\phi$, as well as to
the MSSM flat direction, $\varphi$, which is displaced away from its
minimum (towards large VEVs) during inflation. The governing potential
can be obtained from Eqs.~(\ref{infpot},\ref{chimass})
\beq \label{mssm1}
V = {1 \over 2} m^2_{\phi} {\phi}^2 + g^ 2 {\phi}^2 {\chi}^2 +
{g \over \sqrt{2}} m_{\phi} \phi {\chi}^2 + \lambda^2 {\varphi}^2 {\chi}^2.
\eeq
As mentioned in the previous section, we generically have $\lambda
\geq 3 \times 10^{-4}$, and $g$ can be as large as $\sim {\cal O}(1)$.

After mode decomposition of the field $\chi$, the energy of the mode
with momentum $k$, denoted by $\chi_{k}$, is given by~\cite{preheat2}:
\beq \label{energy}
\omega_k = {\left(k^2 + 2 g^2 {\langle \phi \rangle}^2 +
\sqrt{2} g  m_{\phi}
\langle \phi \rangle +
2 \lambda^2 {\langle \varphi \rangle}^2 \right)}^{1/2}.
\eeq
We have frozen the expansion of the Universe. Including the expansion
will not change our conclusions anyway.  First note that during
inflation the inflaton VEV is large, i.e. $\langle \phi \rangle >
M_{\rm P}$.  Therefore if $g > 10^{-6}$ the inflaton induces a large
mass $g \langle \phi \rangle > H_{\rm I}$ for $\chi$ during inflation.

As a result, $\chi$, quickly settles down to the minimum, i.e.
$\langle \chi \rangle =0$, even if it is initially displaced, and
remains there. Therefore, $\varphi$, does not receive any mass
corrections from its coupling to $\chi$ during inflation.  Note that
the VEV of the flat direction, $\varphi$, induces a large mass,
$\lambda \varphi_0$, to the $\chi$ field during inflation.


At the end of inflation, i.e. when $H(t) \simeq m_{\phi}$, the
inflaton starts oscillating with frequency $m_{\phi}$ and an initial
amplitude ${\cal O}(M_{\rm P})$. Note that for $g > 10^{-6}$ the
quartic inflaton coupling takes over the cubic one in
Eq.~(\ref{mssm1}).
In the interval $m_0 \leq H (t) \leq m_{\phi}$ the flat direction
VEV slides very slowly because of the under damped motion due to
large Hubble friction term, the flat direction effectively slow
rolls. Non-perturbative production of $\chi$ quanta will occur if
there is a non-adiabatic time-variation in the energy, i.e. that ${d
\omega_k/dt} \gs \omega^2_k$. The inflaton oscillations result in a
time-varying contribution to $\omega_k$, while the flat direction
coupling to $\chi$ yields a virtually {\it constant} piece.

Obviously the piece induced by the flat direction VEV weakens the
non-adiabaticity condition~\footnote{Note that in an absence of
$2\lambda^2\langle \varphi\rangle^2$ term in Eq.~(\ref{energy}), the
adiabaticity condition would be violated every time the inflaton
would cross the origin leading to a copious production of $\chi$
particles with momentum $k \ls \left(g {\hat \phi}
m_{\phi}\right)^{1/2}$~\cite{preheat1,preheat2} (${\hat \phi}$ is
the amplitude of the inflaton oscillations).}. Indeed time-variation
of $\omega_k$ will be adiabatic at all times
\beq \label{adia}
{d \omega_k \over dt} < \omega_k^2\,, \eeq
provided that
\begin{equation}
\label{condo0} \lambda^2 \langle \varphi \rangle^2 > g {\hat \phi}
m_{\phi}\,,
\end{equation}
where ${\hat \phi}$ is the amplitude of the inflaton oscillations. We
find the most conservative bound by considering the most optimistic
situation for preheating:
\begin{itemize}
\item
{\it The largest possible amplitude for the inflaton oscillations,
${\hat \phi} \sim {\cal O}(M_P)$.}
\item
{\it The largest possible coupling to the inflaton, $g \sim {\cal
O}(1)$.  }
\end{itemize}
We remind that in the absence of flat direction VEV preheating would
enter an explosive stage when $H(t) \sim 10^{-2} m_{\phi}$, at which time
${\hat \phi} \sim 10^{-2} M_{\rm P}$~\cite{preheat2}.  Therefore, to
prevent efficient preheating, it will be actually sufficient to
satisfy Eq.~(\ref{adia}) at this time rather than the very beginning
of inflaton oscillations. Moreover, as mentioned earlier, having $g
\sim {\cal O}(1)$ also implies a large $\chi$ self-coupling in
supersymmetry. Then one in addition expects preheating to be
considerably suppressed due to self-interactions~\cite{ac,pr}.
Nevertheless we want to find the strongest bound on the flat direction
VEV which shuts-off non-perturbative particle production at {\it all
times} and for the {\it largest coupling} to the inflaton. The natural
conclusion is that there will be no preheating in {\it more} realistic
situations.

It turns out from Eqs.~(\ref{energy},\ref{adia}) that the energy of
mode $\chi_k$ changes adiabatically at all times
if Eq.~(\ref{condo0}) is satisfied at the beginning of inflaton
oscillations, i.e. $H = m_{\phi}$~\footnote{In the absence of
resonant particle production the amplitude of the inflaton
oscillations is redshifted like ${\hat \phi} \propto H(t)$ due to
the Hubble expansion. While, for $H > m_0$, we have $\langle \varphi
\rangle \propto H(t)^{1/(n-2)}$, see Eq.~(\ref{ultivev}). Hence the
RH side of Eq.~(\ref{condo0}) becomes increasingly larger than its
LH side, and
the adiabaticity condition
will be satisfied more comfortably as
time goes by.}. Hence there will be no resonant production of $\chi$
quanta, provided that
\begin{equation} \label{cond}
\varphi_0 >  \lambda^{-1} \left(g M_{\rm P} m_{\phi}\right)^{1/2}.
\end{equation}
This surmounts to a kinematical blocking of preheating by inducing a
piece (which is virtually constant at time scales of interest) to the
mass of inflaton decay products due to their couplings to a flat
direction which has a large VEV.

Once $g {\hat \phi} \sim m_{\phi}$, the Hubble expansion rate becomes:
$H(t) \sim m^2_{\phi}/g M_{\rm P}$, and the cubic interaction term
$\phi \chi^2$ takes over the quartic one. For $m_{\phi} = 10^{13}$~GeV
and $g > 10^{-6}$ this happens when the expansion rate is still $\gg
1$~TeV. Eq.~(\ref{cond}) implies that $\lambda^2 \varphi^2_0 \gg
m^2_{\phi}$, and hence the flat direction VEV totally dominates the
mass of $\chi$ when the cubic term has taken over.  Therefore there
will be no tachyonic instability in the mass of $\chi$ which would
otherwise occur during half of each inflaton oscillation and could lead to
efficient particle production~\cite{dfkpp}.

Now let us find the range of VEVs for which a MSSM flat direction can
satisfy the condition given in Eq.~(\ref{cond}) and shut-off
preheating. We choose a nominal value of the inflaton mass $m_{\phi} =
10^{13}$ GeV. As discussed in the previous subsection, we have
$\lambda \geq 3 \times 10^{-4}$ for all flat directions but few
exceptional ones. The required values of $\varphi_0$ are depicted in
Table.~1 for $\lambda^2 = 10^{-7}$ and $\lambda = 1)^{-1}$. The three
categoric values of inflaton coupling $g = 1,~3 \times
10^{-4},~10^{-6}$ illustrate the distinctive regimes where inflaton
decay would take place in the absence of flat direction VEV:
\begin{itemize}
\item{ $3 \times 10^{-4} \leq g \leq 1$: In this case preheating via
the quartic interaction term $g^2 \phi^2 \chi^2$ would be
efficient~\cite{preheat2}.}
\end{itemize}
\vspace*{5mm}
\begin{center}
\begin{tabular}{|r|r|r|r|}
\hline
  & $g = 1$ & $g = 3 \times 10^{-4}$ & $g = 10^{-6}$ \\
\hline
$\lambda^2 \simeq 10^{-7}$ &  $\varphi_0 \sim M_{\rm P}$ & $\varphi_0 >
3 \times 10^{17}$
& $\varphi_0 > 10^{16}$  \\
\hline
$\lambda^2 \simeq 10^{-1}$ & $\varphi_0 > 10^{16}$ & $\varphi_0
> 3 \times 10^{14}$ & $\varphi_0 > 10^{13}$   \\
\hline
\end{tabular}
\vspace*{2mm}
\end{center}
\noindent
{\bf Table 1:}~ The flat direction VEV, inflaton coupling $g$, and two
values of generic flat direction (Yukawa) couplings $\lambda$ within
MSSM are illustrated. The VEVs are denoted in GeV. For the VEVs above
the quoted numbers preheating is kinematically blocked.
\vspace*{2mm}
\begin{itemize}
\item{ $10^{-6} \leq g < 3 \times 10^{-4}$: In this case preheating
via the quartic term would not be efficient, however, the cubic
interaction term $\left(g m_{\phi}/\sqrt{2}\right) \phi \chi^2$ could
lead to efficient particle production~\cite{dfkpp}.}
\end{itemize}
\begin{itemize}
\item{ $g < 10^{-6}$: In this case inflaton would decay in the
perturbative regime from the beginning, thus no resonant and/or
tachyonic particle production.}
\end{itemize}

It is important to note that preheating is always shut-off for a {\it
sub-Planckian} flat direction VEV. Even in the most extreme case,
i.e. the largest inflaton coupling $g\sim {\cal O}(1)$ and for a flat
direction coupling $\lambda \sim 10^{-4}$, the required VEV for the
flat direction is $\varphi_0 \simeq M_{\rm P}$. Much smaller VEVs,
$\varphi_0\geq 10^{13}$~GeV, are required for moderate values of $g$
and/or $\lambda \sim {\cal O}(1)$ (which is the case for flat
directions including a sizeable component of ${\bf Q}$ and/or ${\bf
u}$ from the third generation).

\subsection{Additional observations}

Some words on fermionic preheating, see Refs.~\cite{Kofman}.
Preheating of superheavy fermions can be much more efficient than
bosonic preheating~\cite{Peloso}.  However within SUSY this is not the
case as the symmetry between bosons and fermions implies similar
equations for the momentum excitations, see Eq.~(\ref{energy}).  The
same also holds for kinematical blocking of preheating by the flat
direction VEV as the induced mass $\lambda \langle \varphi \rangle$,
is SUSY preserving.  Therefore, the condition for the adiabatic change
in vacuum remains as it is in the case of bosonic excitations, and
fermionic preheating is prohibited for the flat direction VEV given in
Eq.~(\ref{cond}).

Finally let us compare the flat direction VEV obtained in the presence
of non-renormalizable terms with those given in Table. 1.  For
$m_{\phi} = 10^{13}$ GeV, and $M = M_{\rm P}$ in Eq.~(\ref{nonren}),
we obtain from Eq.~(\ref{ultivev}) the field values shown in table.~2 right
at the end of inflation (i.e. at $H \simeq m_{\phi}$).
\vspace*{5mm}
\begin{center}
\begin{tabular}{|r|r|r|r|}
\hline
  & $n = 4$ & $n = 6 $ &$n=9$ \\
\hline
$\varphi_0$ &  $ 10^{16}$~GeV & $ 10^{17}$~GeV & $10^{18}$~GeV  \\
\hline
\end{tabular}
\vspace*{2mm}
\end{center}
\noindent
{\bf Table 2:}~ The flat direction VEV at the beginning of inflaton
oscillations in presence of the non-renormalizable term in the
potential, see Eq.~(\ref{nonren}).\\
\vspace*{5mm}

These values comfortably lie within the range depicted in
Table.~1. For $H < m_{\phi}$ the amplitude of inflaton oscillations is
redshifted ${\hat \phi} \propto H$, while the flat direction VEV
slides down to an instantaneous value $\langle \varphi \rangle \sim
\left(H(t) M^{n-3}\right)^{1/n-2}$, see Eq.~(\ref{ultivev}). Therefore
if $\lambda^2 \varphi^2_{0} > g M_{\rm P} m_{\phi}$ at the onset of
inflaton oscillations, we will have $\lambda^2 {\langle \varphi
\rangle}^2 > g {\hat \phi} m_{\phi}$ at later times (note that $n \geq
4$). This implies that even when the flat direction VEV slowly slides
down in a non-renormalizable potential preheating will remain
shut-off.

Once $H(t) \simeq m_0 \sim 1$ TeV, soft SUSY breaking mass term in the
potential takes over and the flat direction starts oscillating around
its origin with an initial amplitude: $\varphi_{\rm in} \sim \left(m_0
M_{\rm P}^{n-3}\right)^{1/n-2}$.  However this happens at time scales
hierarchically longer than those relevant for the preheating
phenomena.

\section{Inflaton life time}

\label{FPI}

Since preheating is kinematically shut-off, the inflaton simply
oscillates with a decreasing amplitude due to the Hubble expansion
rate for $H(t) < m_{\phi}$. The Universe is therefore dominated by the
inflaton oscillations, which for $m^2_{\phi} {\phi}^2$ potential act
as a non-relativistic matter, implying that ${\hat \phi} \propto
H(t)$.

Once $H(t) \simeq m_0$, the flat direction also starts oscillating
and, due to Hubble damping, $\langle \varphi \rangle \propto H$. We
remind that for $g {\hat \phi} \ll m_{\phi}$ the cubic interaction
term $\phi \chi^2$ is dominant and production of $\chi$ can only occur
in a narrow momentum band peaked around $k =
m_{\phi}/2$~\cite{preheat1,preheat2}. Note that at the onset of flat
direction oscillations ${\hat \phi} = m_0 M_{\rm P}/m_{\phi}$, and
hence $g {\hat \phi} \ll m_{\phi}$ even if $g = 1$. Therefore the
inflaton decay will be kinematically forbidden until $\lambda \langle
\varphi \rangle < m_{\phi}/2$.

In the universe which is filled by the inflaton oscillations
$\langle \varphi \rangle \propto H(t)$ for $H < m_0$. Thus the
inflaton eventually decays when the Hubble expansion rate
becomes~\cite{am2}:
\begin{eqnarray} \label{kindec}
H_{\rm d} \simeq {m_{\phi} m_0 \over \lambda \varphi_0} \leq
\left({m_{\phi} \over g M_{\rm P}}\right)^{1/2} m_0 \, .
\end{eqnarray}
Here we have used Eq.~(\ref{cond}) to obtain the inequality on the
RH side. For $m_{\phi} = 10^{13}$ GeV and $m_0 \sim {\cal O}(1)$
TeV, it can be seen that for all couplings in the range $10^{-6}
\leq g \leq 1$ , we have $H_{\rm d} \ll\Gamma_{\rm d}\sim m^3_{\phi}
/g^2 M^2_{\rm P}$, which is equivalent to one-particle decay of the
inflaton. This implies that the inflaton decay becomes kinematically
allowed only in the perturbative regime.

The total decay rate of the inflaton (when we also account for decay
to the fermionic partner of $\chi$) is given by $\Gamma_{\rm d} =
g^2 m_{\phi}/8 \pi$. For $10^{-6} \leq g \leq 1$ we have
$\Gamma_{\rm d} > H_{\rm d}$.  This implies that the inflaton will
immediately decay as soon as kinematics allow~\footnote{The inflaton
decay into two $\chi$ (and its fermionic partner) is kinematically
forbidden for $H > H_{\rm d}$. However higher order decays to light
particles via off-shell $\chi$ (and its fermionic partner) will be
kinematically allowed at all times. The lowest order such process is
four-body (perturbative) inflaton decay via two off-shell $\chi$.
This decay channel, in addition to phase space suppression, is
suppressed by a factor of $\left(m_{\phi}/\varphi_0 \right)^4$. This
results in a very small decay rate which will not be important for
our discussion.}. The above Eq.~(\ref{kindec}) implies~\cite{am2}
\beq \label{gamma}
H_{\rm d} \ls \left(10^{-3}-10^{-1}\right)~{\rm TeV}\,
\eeq
for $m_0 \sim {\cal O}(1)$~TeV.  This underlines the fact that,
regardless of how large its coupling to $\chi$ is, the inflaton will
not decay until after the flat direction has started oscillating, and
even then its decay will be {\it strictly perturbative}.  The
expression in Eq.~(\ref{gamma}) for the inflaton lifetime is very
robust and practically independent from how exactly flat direction
oscillations terminate (see the discussion below).

As pointed out in Refs.~\cite{am1,am2}, when the inflaton completely
decays, the decay products of MSSM fields do not thermalize
promptly. The flat direction VEV also gives masses to gauge bosons and
gauginos which slow down various scattering processes,
i.e. $2\leftrightarrow 2$ and $2\rightarrow 3$, required for a
complete thermalization. The Universe undergoes a bout of
quasi-thermal phase where the plasma obtains (near) kinetic
equilibrium, but not chemical equilibrium. Eventually when the flat
direction starts oscillating and its VEV decreases the Universe
reaches full thermal equilibrium. In a model independent case, the
reheat temperature is governed by the thermalization rate,
i.e. $T_{R}\sim \left(\Gamma_{\rm th}M_{P}\right)^{1/2}$, and can be
as low as TeV~\cite{am2}.  We note that the VEVs required for
kinematical blocking of preheating, see Table. 1, are sufficiently
large to also delay complete thermalization of the
Universe~\cite{am2}.

In some cases the coherent oscillations of the flat direction can
fragment the homogeneous mode into a non-topological
solitons~\cite{Q-ball}, known as Q-balls.  In which case the flat
direction VEV vanishes outside the Q-balls, while still being large
inside them.  Then, since the Q-balls occupy a very tiny fraction of
the space, the inflaton decay will become kinematically allowed as
soon as Q-balls are formed.  In principle flat direction oscillations
might also decay rapidly via preheating due to their gauge
interactions and initial condition $\varphi_0 \gg m_0$~\footnote{This
does not happen for flat directions which have a non-zero A-term
(either from higher-order superpotential or K\"ahler potential
terms). The A-term in this case triggers out-of-phase oscillations of
the real and imaginary parts of the flat direction (with comparable
amplitudes). Then the mass of particles which are coupled to the flat
direction will not experience a non-adiabatic variation, and hence no
preheating~\cite{acs,pm}.}.  An important point is that the time scale
for fragmentation of flat direction oscillations, or their
non-perturbative decay via preheating, lies within the same order as
the RH side of Eq.~(\ref{gamma}). Therefore, irrespective of the fate
of flat direction oscillations, Eq.~(\ref{gamma}) provides a robust
{\it lower bound} on the inflaton lifetime.


\section{Consequences of NO-preheating}
\label{CONS}

The absence of preheating and a delayed perturbative decay of the
inflaton has interesting cosmological consequences which we briefly
discuss here.

\subsection{No gravitino problem}

The foremost consequence of NO preheating is that there will be no
copious production of dangerous relics from scatterings in the aftermath of
inflaton decay. First of all note that there will be no preheat plasma with
large occupation numbers $f_k \geq 1$, which could lead to disastrous
production of relics from scatterings~\cite{am1,kp}. On the other
hand, since the inflaton decays perturbatively, the MSSM fields are
scarcely populated~\cite{am2}.  The challenge is to enhance the number
density of particles which mainly happens via $2 \rightarrow 3$
scatterings. As pointed out in Refs.~\cite{am1,am2} the production of
dangerous relics such as gravitinos is very much suppressed in the
initial plasma.

Large VEVs of the flat directions do not modify the non-thermal
production of gravitinos during the coherent oscillations of the
inflaton~\cite{Maroto,non-pert2,non-pert3}. However this will not be a
threat for BBN. Helicity $\pm 3/2$ are not produced copiously at the
first point, while $ \pm 1/2$ fermions thus produced are mainly
inflatinos, which decay along with the inflaton (thus long before
BBN)~\cite{non-pert3}.

Another source is direct production of gravitinos via the channel
$inflaton \rightarrow inflatino + gravitino$ (if kinematically
open). If the inflaton decay to matter occurs very late, this process
would have a large branching ratio and could lead to overproduction of
gravitinos~\cite{nop,ajm}. However this will not be a problem because
delayed inflaton decay (due to kinematical blocking) occurs at $H_{\rm
d}$, see Eqs.~(\ref{kindec},\ref{gamma}), in which case the direct
gravitino production from inflaton decay is under control.  Similar
analysis holds for moduli fields also since their interaction rates
are also suppressed similar to the case of gravitinos.

\subsection{No non-thermal leptogenesis}

Leptogenesis is a scheme for creating a lepton asymmetry, which is
then partially converted into baryon asymmetry via SM sphalerons, for
a review see~\cite{Buchmuller}. The prospect for thermal leptogenesis
is severely hampered within SUSY due to late
thermalization~\cite{am2}.  This motivates the case for non-thermal
leptogenesis (for example, see~\cite{Rouz-Maz}).

Note that the RH (s)neutrinos can obtain large masses through their
coupling to ${\bf H_u}$ and ${\bf L}$ superfields (for details see
Section~\ref{ICSGS}). If flat directions including ${\bf H}_u$ and/or
${\bf L}$ develop a large VEV then the resonant excitation of the
heavy (s)neutrinos is kinematically forbidden.  If flat directions
which do not include ${\bf H}_u$ or ${\bf L}$ develop a VEV, for
instance ${\bf udd}$, then the RH (s)neutrinos obtain large VEV
dependent masses from the non-renormalizable superpotential through
effective Yukawa couplings~\cite{abm}. In either case the resonant
production of (s)neutrinos is unlikely for the same reason as we
discussed above.  This seriously severs the prospect for non-thermal
leptogenesis from on-shell superheavy (s)neutrinos.


\subsection{No creation of superheavy WIMPS}

In Ref.~\cite{preheat2} it was advocated that it is possible to excite
superheavy a weakly interacting massive particle (WIMP) from
non-perturbative inflaton decay. In most of the cases these WIMPs are
either charged under the SM gauge group, or are a SM gauge singlet
which is coupled to some gauge sector. In either case their production
through non-perturbative decay of the inflaton will be hampered in
SUSY. In the former case a MSSM flat direction developing a VEV
induces a large mass to WIMP which can kinematically block its
production via preheating. The latter case is similar to that of
right-handed (s)neutrinos discussed in the previous subsection.

\subsection{No Exit from hybrid inflation}

So far we have considered the effects of flat directions on preheating
which usually arises in models of chaotic inflation. Here we briefly
discuss the case for models of hybrid inflation.

The simplest SUSY hybrid inflation model has the following
superpotential, see for instance~\cite{Lyth}
\beq \label{hybsuppot}
W \supset y {\bf \Phi} \left({\bf \Psi}^2 - {\Psi^2_0} \right),
\eeq
where ${\bf \Phi}$ is the inflaton superfield and ${\bf \Psi}$ is a
superfield which contains an auxiliary scalar field $\psi$. Inflation
is driven by the false vacuum potential $y^2 {\Psi}^4_0$ during
which $\phi$ undergoes slow-roll motion and $\psi$ is stuck at the
origin.  Inflation ends when $\langle \phi \rangle$ reaches the
critical value $\phi_{\rm c} = {\Psi_0}/\left(\sqrt{2} y\right)$, at which
point the
auxiliary field mass-squared becomes negative and tachyonic preheating
takes place~\cite{tachyonic}. Eventually the two fields settle at $\langle \phi
\rangle= 0$ and $\langle \psi \rangle= {\Psi_0}$.

Hybrid inflation is often quoted as the most successful inflationary
model motivated by particle physics, for a review see~\cite{Lyth}.
However in all cases the inflaton remains a SM gauge singlet. For a
successful phenomenology it is then imperative that the vacuum energy
density during inflation is converted into MSSM particles.

In the simple case given by Eq.~(\ref{hybsuppot}) both the inflaton
and the auxiliary field are SM singlets. Then the ${\bf \Psi}$
superfield can have superpotential couplings to the gauge-invariant
combinations of the MSSM given in Eqs.~(\ref{two},\ref{three}).  Since
$\langle \psi \rangle = 0$ during inflation, any field coupled to
$\psi$ is massless during inflation. This implies that any MSSM flat
direction $\varphi$ which includes a field coupled to ${\psi}$ can
acquire a large VEV, again denoted by $\varphi_0$, in the inflationary
epoch. Such a large VEV will induce a mass $\lambda \varphi_0$ to $\psi$
through the following term in the scalar potential:
\beq \label{hybscalpot}
\lambda^2 {\vert \varphi \vert}^2 \psi^2\,.
\eeq
If $\lambda \varphi_0 > y^{1/2} \Psi_0$, the mass-squared of $\psi$
will remain positive even for $\langle \phi \rangle < \phi_{\rm
c}$. As a result $\langle \psi \rangle = 0$ while $\phi$ is still slow
rolling, and hence tachyonic preheating will never take place. More
importantly it implies that inflation will never end!  Therefore,
unlike chaotic inflation, the absence of preheating in hybrid models
has a negative consequence. Indeed for a graceful exit from the
inflationary phase one needs to have $\lambda \varphi_0 < y^{1/2}
\Psi_0$.

In more realistic models of hybrid inflation the auxiliary field $\psi$
is associated with a Higgs field which spontaneously breaks part(s) of
a larger gauge group which contains the SM, such as grand unified
theory (GUT) or an intermediate scale $U(1)_{B-L}$, upon developing a
VEV after inflation ends (see for instance~\cite{Lyth,Anup-N}). This
Higgs field is naturally coupled to some of the fields which carry
gauge quantum numbers under the larger symmetry. Note that this
symmetry is unbroken during inflation since $\langle \psi \rangle =
0$. Then flat directions which include the fields coupled to ${\psi}$ can
obtain a large VEV during inflation and induce a large mass
to it. Again this can prevent an end to inflation.

Therefore graceful exit from hybrid inflation typically leads to
constraints on the couplings of the auxiliary field to gauge
non-singlets. Note that such couplings are present in any model which
is embedded into a particle physics model. This must be taken into
account in any realistic model of hybrid inflation. We will deal with
some of these issues in a separate publication.

\section{Conclusion}

We argued in this paper that non-perturbative decay of the inflaton
via resonant or tachyonic instabilities is unlikely within SUSY, due to the
presence of flat directions, and
it rather prefers to decay perturbatively.  A flat direction VEV
$> 10^{13}$~GeV will be sufficient to kinematically block preheating.

The key observation is the presence of flat directions. Within MSSM
there are nearly $300$ flat directions, it is expected that a
number (if not all) of independent directions would develop large VEVs
during inflation. The flat directions have Yukawa couplings to the
inflaton decay products and induce large SUSY preserving VEV-dependent
masses to them. For reasonable (sub Planckian) VEVs, depicted in Table.~1,
this leads to a kinematical
blocking of the inflaton decay via preheating, even for ${\cal O}(1)$
inflaton couplings to other fields.

The inflaton decay will be kinematically allowed only after the flat
direction starts oscillating, and once its VEV has been sufficiently
redshifted. We found that the final decay of the inflaton is
perturbative and there exists a robust upper bound on the inflaton
lifetime $\sim (10-10^3) {\rm TeV}^{-1}$.

The absence of a violent stage of non-perturbative inflaton decay has
important implications for particle cosmology. Most notably the
initial plasma has much smaller occupation numbers,
i.e. $f_{k}\ll 1$, which implies that overproduction of dangerous relics
through
scatterings is unlikely, see~\cite{am1,am2,kp}. As we have already
noticed in Ref.~\cite{am2}, the flat direction VEVs suppress the rate
of thermalization, therefore the reheat temperature can be as low as
${\cal O}({\rm TeV})$. In which case thermal production of
gravitinos and other relics will be negligible.

As a cursory remark we pointed out that SUSY hybrid inflation requires
a more careful treatment while taking into account of the flat
directions. Our brief discussion suggests that large VEVs of the flat
directions can even prevent a graceful exit from inflation in these
models.

Finally, since our focus was on the $m^2_{\phi} \phi^2$ case, one
might worry about preheating in $\lambda \phi^4$ model. First of
all, in the pure $\lambda \phi^4$ case preheating is inefficient and
only about $5\%$ of the energy density in inflaton oscillations is
converted into its own quanta~\cite{lambda}.  In any case the $\phi$
quanta must eventually decay to MSSM fields, which will be
kinematically forbidden in the presence of a large flat direction
VEV as we have discussed. Moreover the $\lambda \phi^4$ model is
rather unattractive as
it can produce large non-Gaussianity~\cite{Asko-ng}.


\section{Acknowledgments}

The authors would like to thank Robert Brandenberger and Natalia
Shuhmaher for vigorous discussions and important feedbacks on the
initial draft. We would also like to thank Cliff Burgess, Juan
Garcia-Bellido, Kari Enqvist, Asko Jokinen, Antonio Masiero, Holger
Nielsen, Silvia Pascoli, Leszek Roszkowski and Igor Tkachev for
various discussions.  The work of R.A. is supoported by the Natural
Sciences and Engineering Research Council of Canada (NSERC). A.M.
would like to thank CERN, University of Padova, Perimeter Institute
and McGill University for their kind hospitality during the course
of this project.


\section{Appendix}

\subsection{MSSM flat directions}\label{BRM}

There exist a large number of directions in the field space of
supersymmetric theories, known as flat directions, along which the
scalar potential identically vanishes in the limit of exact SUSY.  In
this limit the scalar potential of MSSM, denoted by $V_{\rm MSSM}$, is
the sum of the $F$- and $D$-terms and reads
\begin{equation}
\label{fplusd}
V= \sum_i |F_i|^2+\frac 12 \sum_a g_a^2D^aD^a\,,
\end{equation}
where
\begin{equation}
F_i \equiv {\partial W_{\rm MSSM}\over \partial \chi_i},~~D^a=\chi^\ast_i
T^a_{ij}\chi_j~\,.
\label{fddefs}
\end{equation}
Here the scalar fields, denoted by $\chi_i$, transform under a gauge
group $G$ with the generators of the Lie algebra and gauge coupling
are by $T^{a}$ and $g_a$ respectively.

For a general supersymmetric model with $N$ chiral superfields, it is
possible to find out the directions along which the potential in
Eq.~(\ref{fplusd}) vanishes identically by solving simultaneously
\begin{equation}
\label{fflatdflat}
\chi^{\ast}_i T^{a}_{ij} \chi_j = 0\,, \quad \quad
\frac{\partial W}{\partial \chi_{i}}=0\,.
\end{equation}
Field configurations obeying Eq.~(\ref{fflatdflat}) are called
respectively $D$-flat and $F$-flat.

$D$-flat directions are parameterized by gauge-invariant monomials of
the chiral superfields. A powerful tool for finding the flat
directions has been developed
in~\cite{buccella82,affleck84,ad,drt,luty96,gkm}, where the
correspondence between gauge invariance and flat directions has been
employed. In particular all flat directions have been classified
within MSSM~\cite{gkm}.

Adding the inflaton superfield which is a SM singlet does not affect
$D$-flatness. However one might worry that the inflaton coupling to
matter would ruin the $F$-flatness as new terms can now appear in the
superpotential. For example consider the non-renormalizable
superpotential terms:
\begin{eqnarray} \label{supcoup}
{1 \over M_{\rm P}} {\bf \Phi} {\bf H_u} {\bf Q} {\bf u}, ~ {1 \over M_{\rm P}}
{\bf \Phi} {\bf H_d}
{\bf Q} {\bf d}, ~ {1 \over M_{\rm P}} {\bf \Phi} {\bf H_d}
{\bf L} {\bf e} \, \nonumber \\
{1 \over M_{\rm P}} {\bf \Phi} {\bf Q} {\bf L} {\bf d}, ~ {1 \over M_{\rm P}}
{\bf \Phi} {\bf u} {\bf d} {\bf d}, ~
{1 \over M_{\rm P}} {\bf \Phi} {\bf L} {\bf L} {\bf e} \,.
\end{eqnarray}
which can arise in addition to the renormalizable one in
Eq.~(\ref{rensup}). If ${\Psi} = {\bf H}_d$, see Eq.~(\ref{infsup}),
terms in the first row preserve $R$-parity while those in the second
row violate it. The reverse situation happens if ${\bf \Psi} = {\bf
L}$. If $R$-parity is a discrete subgroup of a gauge symmetry, it
will remain unbroken by gravitational effects. In this case only
those terms in Eq.~(\ref{supcoup}) which preserve $R$-parity can
appear in the superpotential. On the other hand, like other global
symmetries, $R$-parity will be supposedly broken due to
gravitational effects if it is not protected by some gauge symmetry
(see for example Ref.~\cite{lenny}). In this case the superpotential
can include all terms in Eq.~(\ref{supcoup}) regardless of what
${\bf \Psi}$ represents.

Further note that in this case the LSP decays via terms which violate
$R$-parity. However the decay is suppressed by both $M_{\rm P}$ and
$m_{\phi}$, and hence the LSP lifetime is much longer than the age of
the Universe. These interactions can also led to proton decay though
at time scale which is much longer than the experimental
bound. Therefore $R$-parity breaking through $M_{\rm P}$ suppressed
operators will not be constrained.

In models of large field inflation, in which $\langle \phi \rangle >
M_{\rm P}$ during inflation,
the terms in Eq.~(\ref{supcoup}) result in effective renormalizable
superpotential terms. Note that terms from the first row are exactly
the same as the SM Yukawa couplings, and hence do not lead to any new
constraints.  On the other hand, the second row results in terms which
are absent in the MSSM superpotential. Therefore, if allowed, they
will lead to new $F$-flatness constraints and lift some of the flat
directions.

Note however that a large subset of MSSM flat directions survive. In
particular those which are only made up of ${\bf Q},{\bf u},{\bf e}$
superfields will {\it not be affected at all} . Interestingly these
directions are coupled through the MSSM superpotential in
Eq.~(\ref{mssm}) to {\it both} of the ${\bf H}_u$ and ${\bf \Psi}$
superfields, whether ${\bf \Psi} = {\bf H}_d$ or ${\bf \Psi} = {\bf
L}$, see Eq.~(\ref{infsup}). As we will see, this is very important
for shutting off preheating.

\subsection{Inflaton couplings to SM gauge singlets}
\label{ICSGS}

Inflaton being a gauge singlet need not directly couple to the (MS)SM
sector, but instead can do so through another SM gauge singlet.
Phenomenologically the best motivated example is when the inflaton is
coupled to the RH (s)neutrinos. Note that the couplings
of RH (s)neutrinos to the (MS)SM sector can explain the origin of
light neutrino masses via see-saw mechanism~\cite{see-saw}.

Assuming that the RH (s)neutrinos obtain masses from some other
source, the relevant part of the superpotential will be given by:
\beq
\label{super}
W \supset {1 \over 2} m_\phi {\bf \Phi} {\bf \Phi} + {1 \over 2} g
{\bf \Phi} {\bf N}{\bf N} + h {\bf N} {\bf H}_u {\bf L} + {1 \over 2} M_N
{\bf N} {\bf N}\,.
\eeq
Here ${\bf \Phi},{\bf N},{\bf L},{\bf H}_u$ stand for the
inflaton, the RH neutrino, the lepton doublet, and the Higgs (which
gives mass to the top quark) superfields, respectively.  Also,
$m_\phi$ and $M_N$ denote inflaton and RH (s)neutrino masses,
respectively. Here $h$ denotes the $3 \times 3$ neutrino Yukawa
matrix. For simplicity, we have omitted all indices in $h$ matrix and
superfields, and work in the basis where $M_N$ is diagonal.  The
inflaton coupling to the RH (s)neutrinos can be quite large: $g \sim
{\cal O}(1)$.

Any flat direction that includes ${\bf H}_u$ and/or ${\bf
L}$ can induce a large mass to ${\bf N}$ through renormalizable
couplings.
Moreover note that ${\bf N}$ can also couple to MSSM fields via
non-renormalizable superpotential terms
%
%
the same as Eq.~(\ref{supcoup}) with ${\bf \Phi}$ replaced by
${\bf N}$.  In the presence of large flat direction VEVs,
non-renormalizable interactions in Eq.~(\ref{supcoup}) can lead to
large effective couplings between ${\bf N}$ and MSSM
fields~\cite{abm}.  For a reasonable flat direction VEV, one can have
$\lambda_{\rm eff} \geq 3 \times 10^{-4}$. Then non-perturbative decay
of the inflaton to RH (s)neutrinos will be kinematically forbidden
similar to our earlier analysis as in Section~\ref{FDCIDP}.

\subsection{Categorizing flat directions}\label{append}

A close inspection to the MSSM flat directions shows that they belong
to one of the following groups (for instance see~\cite{Enqvist-REV}):
\begin{itemize}
\item[1-] {Directions which include two or more ${\bf Q}$ from
different generations: ${\bf Q} {\bf Q} {\bf Q} {\bf L}$. These
directions have Yukawa couplings $\geq 10^{-3}$ to both ${\bf H}_u$ and
${\bf H}_d$ both.
}
\item[2-] {Directions which includes two or more ${\bf u}$ from
different generations: ${\bf u} {\bf u} {\bf d} {\bf e}$ and ${\bf u}
{\bf u} {\bf u} {\bf e} {\bf e}$. These directions have Yukawa couplings
$\geq 10^{-3}$ to ${\bf H}_u$.
}
\item[3-] {Directions which include two or more ${\bf d}$ from
different generations: ${\bf u} {\bf d} {\bf d}$ and ${\bf L} {\bf L}
{\bf d} {\bf d} {\bf d}$ . These directions have Yukawa couplings $\geq
10^{-3}$ to ${\bf H}_d$.
}
\item[4-] {Directions which include ${\bf Q}$ and ${\bf u}$ from
different generations: ${\bf Q} {\bf u} {\bf L} {\bf e}$ and ${\bf Q}
{\bf u} {\bf Q} {\bf u} {\bf e}$ . These directions have a coupling
$\geq 10^{-3}$ to ${\bf H}_u$.
}
\item[5-] {Directions which include ${\bf Q}$ and ${\bf d}$ from
different generations: ${\bf Q} {\bf d} {\bf L}$. These directions
have Yukawa couplings $\geq 10^{-3}$ to ${\bf H}_d$.}
\item[6-] {Directions which include two ${\bf L}$ from different
generations: ${\bf L} {\bf L} {\bf e}$. These directions have Yukawa
couplings $\geq 10^{-3}$ to ${\bf H}_d$.
}
\end{itemize}
Note that the requirement that more than one generation of squarks
and/or sleptons be involved comes as a direct consequence of F- and
D-flatness~\cite{gkm}. Now let us consider the ${\bf \Psi} = {\bf
H}_d$ and ${\bf \Psi} = {\bf L}$ cases separately, see
Eqs.~(\ref{infsup},\ref{def1}).
\vskip20pt
\noindent

{${\bf \Psi} = {\bf H}_d$. In this case $\lambda_1 \geq 10^{-3}$
and/or $\lambda_2 \geq 10^{-3}$ for {\it all} of the MSSM flat
directions, see Eq.~(\ref{rel}), implying that $\lambda \geq 3 \times
10^{-4}$.}
\vskip20pt
\noindent
{${\bf \Psi} = {\bf L}$. In this case $\lambda_1 \geq 10^{-3}$, hence
$\lambda \geq 3 \times 10^{-4}$, for flat directions listed in
$1,~2,~4$. There are exceptional flat directions for which $\lambda
\ll 10^{-3}$:
\begin{itemize}
\item{ ${\bf u} {\bf d} {\bf d}$: there are three such directions
(with ${\bf u}$ belonging to the first generation) for which
$\lambda_1 \sim {\cal O} (10^{-5})$ and $\lambda_2 = 0$, hence
$\lambda \sim {\cal O}(10^{-5})$.  }
\item{ ${\bf Q} {\bf d} {\bf L}$: there are 6 such directions (with
${\bf Q}$ belonging to the first generation) for which $\lambda_1 \sim
{\cal O}(10^{-5}$ and $\lambda_2 = 0$, hence $\lambda \sim {\cal
O}(10^{-5})$.  }
\item{ ${\bf L} {\bf L} {\bf d} {\bf d} {\bf d}$: there is one such
direction (where the two ${\bf L}$ are orthogonal to ${\bf \Psi}$. As
mentioned earlier, $\psi$ cannot acquire a large VEV since its
coupling to the inflaton induces a mass $\gg H_{\rm I}$.) for which
$\lambda_1 = \lambda_2 = 0$, hence $\lambda = 0$.  }
\item{ ${\bf L} {\bf L} {\bf e}$: there is one such direction (where
the two ${\bf L}$ are orthogonal to ${\bf \Psi}$) for which $\lambda_1
= 0$ and $\lambda_2 \sim {\cal O}(10^{-5})$, hence $\lambda \sim {\cal
O}(10^{-5})$.  }
\end{itemize}

The exceptional directions constitute a small subset of all MSSM flat
directions: $11$ out of nearly $300$. Moreover, considering
multi-dimensionality of the space of all flat directions, it is very
unlikely that a VEV grows exactly along one of the exceptional
directions during inflation. To elucidate consider flat directions
represented by the ${\bf u} {\bf d} {\bf d}$ monomial. When different
generations are taken into account, this monomial represents a space
of complex dimension $9$~\cite{gkm}. While, exceptional directions
span a three-dimensional subspace. Therefore even with a probabilistic
argument it is very hard to imagine that a non-zero VEV will be
confined to this subspace of flat directions. For the bulk of
$9$-dimensional space the superfield, ${\bf u}$, has comparable
components from all generations, implying that $\lambda_1 \gs 10^{-3}$
and $\lambda \geq 3 \times 10^{-4}$.




\begin{thebibliography}{99}


\bibitem{am1}
R.~Allahverdi and A.~Mazumdar,
  ``Quasi-thermal universe and its implications for gravitino production,
baryogenesis and dark matter,''
  arXiv:hep-ph/0505050.

\bibitem{am2}
R.~Allahverdi and A.~Mazumdar,
  ``Supersymmetric thermalization and quasi-thermal universe: Consequences for
  gravitinos and leptogenesis,''
  arXiv:hep-ph/0512227.


\bibitem{Lyth}
 D.~H.~Lyth and A.~Riotto,
  ``Particle physics models of inflation and the cosmological density
  perturbation,''
  Phys.\ Rept.\  {\bf 314}, 1 (1999)
  [arXiv:hep-ph/9807278].


\bibitem{BBN}
For a review, see: K. A. Olive, G. Steigman and T. P. Walker,
Phys. Rept. {\bf 333}, 389 (2000).

\bibitem{reheat}
A. Dolgov and A. D. Linde, Phys. Lett. B {\bf 116}, 329 (1982);
L. F. Abbott, E. Farhi and M. Wise, Phys. Lett. B {\bf 117}, 29 (1982).
%


\bibitem{Thermalization}
S. Davidson and S. Sarkar, JHEP {\bf 0011}; 012 (2000) [arXiv:hep-ph/0009078].
 R.~Allahverdi,
  Phys.\ Rev.\ D {\bf 62}, 063509 (2000)
  [arXiv:hep-ph/0004035].
  R. Allahverdi and M. Drees, Phys. Rev. D {\bf 66},
063513 (2002) [arXiv:hep-ph/0205246].

\bibitem{Jaikumar}
P. Jaikumar and A. Mazumdar, Nucl. Phys. B {\bf 683},
264 (2004) [arXiv:hep-ph/0212265].


\bibitem{preheat1}
J. Traschen and R. Brandenberger, Phys. Rev. D {\bf 42}, 2491 (1990).
%
\bibitem{preheat2}
L. Kofman, A. D. Linde and A. A. Starobinsky, Phys. Rev. Lett. {\bf
73}, 3195 (1994) [arXiv:hep-ph/9405187];
L. Kofman, A. D. Linde and A. A. Starobinsky, Phys. Rev. D {\bf 56},
3258 (1997) [arXiv:hep-ph/9704452];
Y.~Shtanov, J.~H.~Traschen and R.~H.~Brandenberger,
Phys.\ Rev.\ D {\bf 51}, 5438 (1995) [arXiv:hep-ph/9407247];
 D.~Boyanovsky, H.~J.~de Vega and R.~Holman,
 arXiv:hep-ph/9701304;
 D.~Cormier, K.~Heitmann and A.~Mazumdar,
  Phys.\ Rev.\ D {\bf 65}, 083521 (2002)
  [arXiv:hep-ph/0105236].

\bibitem{Enqvist-Soliton}
 K.~Enqvist, S.~Kasuya and A.~Mazumdar,
  Phys.\ Rev.\ Lett.\  {\bf 89}, 091301 (2002)
  [arXiv:hep-ph/0204270].
 K.~Enqvist, S.~Kasuya and A.~Mazumdar,
  Phys.\ Rev.\ D {\bf 66}, 043505 (2002)
  [arXiv:hep-ph/0206272].


\bibitem{Jed}
For example, see: K. Jedamzik, Class. Quant. Grav. {\bf 19},
3417 (2002) [arXiv:astro-ph/0112226].
%

\bibitem{Maroto}
A.~L.~Maroto and A.~Mazumdar,
Phys. Rev. Lett. {\bf 84}, 1655 (2000) [arXiv:hep-ph/9904206].

\bibitem{non-pert2}
R.~Kallosh, L.~Kofman, A.~D.~Linde and A.~Van Proeyen,
Phys. Rev. D {\bf 61}, 103503 (2000) [arXiv:hep-th/9907124];
R.~Kallosh, L.~Kofman, A.~D.~Linde and A.~Van Proeyen,
Class. Quant. Grav.  {\bf 17}, 4269 (2000) [arXiv:hep-th/0006179];
A. L. Maroto and J. R. Pelaez, Phys. Rev. D {\bf 62},
023518 (2000) [arXiv:hep-ph/9912212];
G.~F.~Giudice, A.~Riotto and I.~I.~Tkachev,
JHEP {\bf 9911}, 036 (1999) [arXiv:hep-ph/9911302];
G.~F.~Giudice, I.~I.~Tkachev and A.~Riotto,
JHEP {\bf 9908}, 009 (1999) [arXiv:hep-ph/9907510];
M.~Bastero-Gil and A.~Mazumdar,
Phys. Rev. D {\bf 62}, 083510 (2000) [arXiv:hep-ph/0002004].
%
\bibitem{non-pert3}
R.~Allahverdi, M.~Bastero-Gil and A.~Mazumdar, Phys. Rev. D {\bf 64}, 023516
(2001) [arXiv:hep-ph/0012057].
H.~P.~Nilles, M.~Peloso and L.~Sorbo,
Phys. Rev. Lett. {\bf 87}, 051302 (2001) [arXiv:hep-ph/0102264].
H.~P.~Nilles, M.~Peloso and L.~Sorbo,
JHEP {\bf 0104}, 004 (2001) [arXiv:hep-th/0103202].
%

\bibitem{Moduli}
G.~F.~Giudice, A.~Riotto and I.~I.~Tkachev,
  JHEP {\bf 0106}, 020 (2001)
  [arXiv:hep-ph/0103248].



\bibitem{Robert-Iso}
F.~Finelli and R.~H.~Brandenberger,
  Phys.\ Rev.\ Lett.\  {\bf 82}, 1362 (1999)
  [arXiv:hep-ph/9809490].
F.~Finelli and R.~H.~Brandenberger,
  Phys.\ Rev.\ D {\bf 62}, 083502 (2000)
  [arXiv:hep-ph/0003172].

\bibitem{Igor}
S.~Y.~Khlebnikov and I.~I.~Tkachev,
  Phys.\ Rev.\ D {\bf 56}, 653 (1997)
  [arXiv:hep-ph/9701423].

\bibitem{Enqvist-NG}
K.~Enqvist, A.~Jokinen, A.~Mazumdar, T.~Multamaki and A.~Vaihkonen,
  Phys.\ Rev.\ Lett.\  {\bf 94}, 161301 (2005)
  [arXiv:astro-ph/0411394].
 K.~Enqvist, A.~Jokinen, A.~Mazumdar, T.~Multamaki and A.~Vaihkonen,
  JCAP {\bf 0503}, 010 (2005)
  [arXiv:hep-ph/0501076].
K.~Enqvist, A.~Jokinen, A.~Mazumdar, T.~Multamaki and A.~Vaihkonen,
  JHEP {\bf 0508}, 084 (2005)
  [arXiv:hep-th/0502185].
N.~Barnaby and J.~M.~Cline,
  Phys. Rev. D {\bf 73}, 106012 (2006) [arXiv:astro-ph/0601481].
N. Barnaby and J. M. Cline, Phys. Rev. D {\bf 75}, 086004 (2007)
[arXiv:astro-ph/0611750].


\bibitem{Asko-ng}
A.~Jokinen and A.~Mazumdar,
  arXiv:astro-ph/0512368.

\bibitem{Peloso}
G. F. Giudice, M. Peloso, A. Riotto and I. I. Tkachev, JHEP {\bf 9908}, 014
(1999) [arXiv:hep-ph/9905242].



\bibitem{Tkachev}
 I.~I.~Tkachev,
  Phys.\ Lett.\ B {\bf 376}, 35 (1996)
  [arXiv:hep-th/9510146].
S.~Khlebnikov, L.~Kofman, A.~D.~Linde and I.~Tkachev,
  Phys.\ Rev.\ Lett.\  {\bf 81}, 2012 (1998)
  [arXiv:hep-ph/9804425].
 I.~Tkachev, S.~Khlebnikov, L.~Kofman and A.~D.~Linde,
  Phys.\ Lett.\ B {\bf 440}, 262 (1998)
  [arXiv:hep-ph/9805209].



\bibitem{gkm}
T. Gherghetta, C. Kolda and S. P. Martin, Nucl. Phys. B {\bf 468},
37 (1996) [arXiv:hep-ph/9510370].

\bibitem{ASKO}
 K.~Enqvist, A.~Jokinen and A.~Mazumdar,
  JCAP {\bf 0401}, 008 (2004)
  [arXiv:hep-ph/0311336].

\bibitem{Enqvist-REV}
 K. Enqvist and A. Mazumdar, Phys. Rept. {\bf 380}, 99
(2003) [arXiv:hep-ph/0209244].
M. Dine and A. Kusenko, Rev. Mod. Phys. {\bf 76}, 1 (2004)
[arXiv:hep-ph/0303065].

\bibitem{ejm}
 K.~Enqvist, A.~Jokinen and A.~Mazumdar,
  JCAP {\bf 0411}, 001 (2004)
  [arXiv:hep-ph/0404269].

\bibitem{fk}
G.~N.~Felder and L.~Kofman,
  Phys.\ Rev.\ D {\bf 63}, 103503 (2001)
  [arXiv:hep-ph/0011160].
%
\bibitem{mt}
R.~Micha and I.~I.~Tkachev,
  arXiv:hep-ph/0301249.
R.~Micha and I.~I.~Tkachev,
  Phys.\ Rev.\ D {\bf 70}, 043538 (2004)
  [arXiv:hep-ph/0403101].

%
\bibitem{kp}
D.~I.~Podolsky, G.~N.~Felder, L.~Kofman and M.~Peloso,
  Phys.\ Rev.\ D {\bf 73}, 023501 (2006)
  [arXiv:hep-ph/0507096].

%
\bibitem{Robert-REV}
V.~F.~Mukhanov, H.~A.~Feldman and R.~H.~Brandenberger,
  Phys.\ Rept.\  {\bf 215}, 203 (1992).
%

\bibitem{Natalia}
N.~Shuhmaher and R.~Brandenberger,
  Phys.\ Rev.\ D {\bf 73}, 043519 (2006)
  [arXiv:hep-th/0507103].

%
\bibitem{dfkpp}
 J.~F.~Dufaux, G.~Felder, L.~Kofman, M.~Peloso and D.~Podolsky,
  arXiv:hep-ph/0602144.


%

\bibitem{abm}
 R.~Allahverdi, R.~Brandenberger and A.~Mazumdar,
Phys.\ Rev.\ D {\bf  70}, 083535 (2004) [arXiv:hep-ph/0407230].

\bibitem{marieke}
K.~Enqvist, A.~Mazumdar and M.~Postma,
Phys. Rev. D {\bf 67}, 121303 (2003) [arXiv:astro-ph/0304187];
A. Mazumdar and M. Postma,
Phys. Lett. B {\bf 573}, 5 (2003) [Erratum-ibid. B {\bf 585}, 295 (2004)]
[arXiv:astro-ph/0306509].
R. Allahverdi, Phys. Rev. D {\bf 70}, 043507 (2004) [arXiv:astro-ph/0403351].

\bibitem{Asm}
 R.~Brandenberger, P.~M.~Ho and H.~c.~Kao,
  JCAP {\bf 0411}, 011 (2004)
  [arXiv:hep-th/0312288].
 A.~Jokinen and A.~Mazumdar,
  Phys.\ Lett.\ B {\bf 597}, 222 (2004)
  [arXiv:hep-th/0406074].


\bibitem{Liddle}
A.~R.~Liddle, A.~Mazumdar and F.~E.~Schunck,
 Phys.\ Rev.\ D {\bf 58}, 061301 (1998)
[arXiv:astro-ph/9804177];
E.~J.~Copeland, A.~Mazumdar and N.~J.~Nunes,
Phys.\ Rev.\ D {\bf 60}, 083506 (1999)
[arXiv:astro-ph/9904309];
A.~Mazumdar, S.~Panda and A.~Perez-Lorenzana,
Nucl.\ Phys.\ B {\bf 614}, 101 (2001)  [arXiv:hep-ph/0107058].


\bibitem{sninfl}
H. Murayama, H. Suzuki, T. Yanagida and J. Yokoyama, Phys. Rev. Lett.
{\bf 70}, 1912 (1993); H. Murayama, H. Suzuki, T. Yanagida and J. Yokoyama,
Phys. Rev. D {\bf 50}, 2356 (1994) [arXiv:hep-ph/9311326].


\bibitem{ac}
  R.~Allahverdi and B.~A.~Campbell,
  Phys.\ Lett.\ B {\bf 395}, 169 (1997)
  [arXiv:hep-ph/9606463].


%
\bibitem{pr}
T.~Prokopec and T.~G.~Roos,
  Phys.\ Rev.\ D {\bf 55}, 3768 (1997)
  [arXiv:hep-ph/9610400].


\bibitem{drt}
M.~Dine, L.~Randall and S.~Thomas, Phys. Rev. Lett. {\bf 75}, 398 (1995)
[arXiv:hep-ph/9503303].
M. Dine, L. Randall and S. Thomas, Nucl. Phys. B {\bf 458}, 291 (1996)
[arXiv:hep-ph/9507453].

\bibitem{NO-Scale}
 A.~B.~Lahanas and D.~V.~Nanopoulos,
  Phys.\ Rept.\  {\bf 145}, 1 (1987).

\bibitem{gmo}
M.~K.~Gaillard, H.~Murayama and K.~A.~Olive, Phys. Lett. B {\bf 355}, 71
(1995) [arXiv:hep-ph/9504307].
%

\bibitem{dis}
M. Dine and N. Seiberg, Nucl. Phys. B {\bf 160}, 243 (1985).


\bibitem{adm}
R. Allahverdi, M. Drees and A. Mazumdar, Phys. Rev. D {\bf 65},
065010 (2002) [arXiv:hep-ph/0108225].

\bibitem{Kofman}
P.~B.~Greene and L.~Kofman,
  Phys.\ Lett.\ B {\bf 448}, 6 (1999)
  [arXiv:hep-ph/9807339].
 J.~Baacke, K.~Heitmann and C.~Patzold,
  Phys.\ Rev.\ D {\bf 58}, 125013 (1998)
  [arXiv:hep-ph/9806205].
A.~L.~Maroto and A.~Mazumdar,
  Phys.\ Rev.\ D {\bf 59}, 083510 (1999)
  [arXiv:hep-ph/9811288].


\bibitem{Q-ball}
A.~Kusenko,
  Phys.\ Lett.\ B {\bf 405}, 108 (1997)
  [arXiv:hep-ph/9704273].
  A.~Kusenko,
  Phys.\ Lett.\ B {\bf 404}, 285 (1997)
  [arXiv:hep-th/9704073].
A.~Kusenko and M.~E.~Shaposhnikov,
  Phys.\ Lett.\ B {\bf 418}, 46 (1998)
  [arXiv:hep-ph/9709492].
 K.~Enqvist and J.~McDonald,
  Phys.\ Lett.\ B {\bf 425}, 309 (1998)
  [arXiv:hep-ph/9711514].
 K.~Enqvist and J.~McDonald,
  Nucl.\ Phys.\ B {\bf 538}, 321 (1999)
  [arXiv:hep-ph/9803380].

\bibitem{acs}
 R.~Allahverdi, R.~H.~A.~Shaw and B.~A.~Campbell,
  Phys.\ Lett.\ B {\bf 473}, 246 (2000)
  [arXiv:hep-ph/9909256].

\bibitem{pm}
 M.~Postma and A.~Mazumdar,
  JCAP {\bf 0401}, 005 (2004)
  [arXiv:hep-ph/0304246].

\bibitem{nop}
H. P. Nilles, K. A. Olive and M. Peloso,
Phys. Lett. B {\bf 522}, 304 (2001) [arXiv:hep-ph/0107212].
%
\bibitem{ajm}
R. Allahverdi, A. Jokinen and A. Mazumdar,
Phys. Rev. D {\bf 71}, 043505 (2005) [arXiv:hep-ph/0410169].
R.~Allahverdi, S.~Hannestad, A.~Jokinen, A.~Mazumdar and S.~Pascoli,
  arXiv:hep-ph/0504102.


\bibitem{Buchmuller}
W. Buchm\"uller, R. D. Peccei and T. Yanagida, arXiv:hep-ph/0502169.
%

\bibitem{Anup-N}
  A.~Mazumdar,
  Phys.\ Lett.\ B {\bf 580}, 7 (2004)
  [arXiv:hep-ph/0308020].

\bibitem{see-saw}
P.~Minkowski, Phys. Lett. B {\bf 67}, 421 (1977).
M. Gell-Mann, P. Ramond and R. Slansky, in {\it Supergravity},
eds. P. van Nieuwenhuizen and D. Z. Freedman (North Holland, 1979).
T. Yanagida, Proceedings of {\it Workshop on
Unified Theory and Baryon number in the Universe}, eds.
O. Sawada and A. Sugamoto (KEK, Tsukuba, 1979).
R. N. Mohapatra and G. Senjanovic, Phys. Rev. Lett. {\bf 44}, 912 (1980).

\bibitem{Rouz-Maz}
G. Lazarides and Q. Shafi, Phys. Lett. B {\bf 258}, 305 (1991).
R. Allahverdi and A. Mazumdar, Phys. Rev. D {\bf 67}, 023509 (2003)
[arXiv:hep-ph/0208268].
T. Dent, G. Lazarides and R. Ruiz de Austri, Phys. Rev. D {\bf 69},
075012 (2004) [arXiv:hep-ph/0312033].
T. Dent, G. Lazarides and R. Ruiz de Austri, arXiv:hep-ph/0503235.
A.~Mazumdar,
Phys.\ Rev.\ Lett.\  {\bf 92}, 241301 (2004)
[arXiv:hep-ph/0306026].
R. Allahverdi, B. Dutta and A. Mazumdar,
Phys. Rev. D {\bf 67}, 123515 (2003) [arXiv:hep-ph/0301184].
H. Murayama and T. Yanagida, Phys. Lett. B {\bf 322}, 349 (1994)
[arXiv:hep-ph/9310297].
K. Hamaguchi, H. Murayama and T. Yanagida, Phys. Rev. D {\bf 65},
043512 (2002) [arXiv:hep-ph/0109030].
Z. Berezhiani, A. Mazumdar and A. P\'erez-Lorenzana,
Phys. Lett. B {\bf 518}, 282 (2001) [arXiv:hep-ph/0107239].
R. Allahverdi and M. Drees, Phys. Rev. D {\bf 69}, 103522 (2004)
[arXiv:hep-ph/0401054].
R. Allahverdi and M. Drees, Phys. Rev. D {\bf 70}, 123522 (2004)
[arXiv:hep-ph/0408289];
%

\bibitem{bdps}
L. Boubekeur, S. Davidson, M. Peloso and L. Sorbo, Phys. Rev. D {\bf 67},
043515 (2003) [arXiv:hep-ph/0209256].


\bibitem{tachyonic}

G. N. Felder, J. Garcia-Bellido, P. B. Greene, L. Kofman, A. D. Linde and I.
I. Tkachev, Phys. Rev. Lett. {\bf 87}, 011601 (2001) [arXiv:hep-ph/0012142].
G. N. Felder, L. Kofman and A. D. Linde, Phys. Rev. D {\bf 64}, 123517 (2001)
[arXiv:hep-ph/0106179].


\bibitem{lambda}

P. B. Greene, L. Kofman, A. D. Linde and A. A. Starobinsky, Phys. Rev. D
{\bf 56}, 6175 (1997) [arXiv:hep-ph/9705347].


\bibitem{lenny}

R. Kallosh, A. D. Linde, D. A. Linde and L. Susskind, Phys. Rev. D
{\bf 52}, 912 (1995) [arXiv:hep-th/9502069].

\bibitem{buccella82}
F. Buccella, J. P. Derendinger, S. Ferrara, and C. A. Savoy, Phys.
Lett. B {\bf 115}, 375 (1982).
%
\bibitem{affleck84}
I. Affleck, M. Dine, and N. Seiberg, Nucl. Phys. B {\bf 241}, 493
(1984).
%
\bibitem{ad}
I. Affleck, M. Dine, and N. Seiberg, Nucl. Phys. B {\bf 256}, 557
(1985).


\bibitem{luty96}
M. A. Luty, and W. Taylor, Phys. Rev. D {\bf 53}, 3399 (1996).
[arXiv:hep-th/9506098].
%




\end{thebibliography}
\end{document}